# The acceleration of cosmic-ray protons in the supernova remnant RX J1713.7-3946

R. Enomoto*, T. Tanimori†, T. Naito‡, T. Yoshida§, S. Yanagita§, M. Mori*, P.G. Edwards∥, A. Asahara†, G.V. Bicknell¶, S. Gunji#, S. Hara**, T. Hara‡, S. Hayashi††, C. Itoh§, S. Kabuki*, F. Kajino††, H. Katagiri*, J. Kataoka†, A. Kawachi*, T. Kifune‡‡, H. Kubo†, J. Kushida**, S. Maeda††, A. Maeshiro††, Y. Matsubara§§, Y. Mizumoto∥∥, M. Moriya**, H. Muraishi¶¶, Y. Muraki§§, T. Nakase##, K. Nishijima##, M. Ohishi*, K. Okumura*, J.R. Patterson***, K. Sakurazawa**, R. Suzuki*, D.L. Swaby***, K. Takano**, T. Takano#, F. Tokanai#, K. Tsuchiya*, H. Tsunoo*, K. Uruma##, A. Watanabe#, and T. Yoshikoshi†††

*ICRR, Univ. of Tokyo, Kashiwa, Chiba 277-8582, Japan
†Kyoto University, Sakyo-ku, Kyoto 606-8502, Japan
‡Yamanashi Gakuin University, Kofu, Yamanashi 400-8575, Japan
§Ibaraki University, Mito, Ibaraki 310-8512, Japan
∥ISAS, Sagamihara, Kanagawa 229-8510, Japan
¶MSSSO, Australian National University, ACT 2611, Australia
#Yamagata University, Yamagata, Yamagata 900-8560, Japan
**Tokyo Institute of Technology, Meguro-ku, Tokyo 152-8551, Japan
††Konan University, Kobe, Hyogo 658-8501, Japan
‡‡Shinshu University, Nagano, Nagano 380-8553, Japan
§§STE Laboratory, Nagoya University, Nagoya, Aichi 464-8601, Japan
∥∥NAO of Japan, Mitaka, Tokyo 181-8588, Japan
¶¶Ibaraki Prefectural University of Health Sciences, Ami, Ibaraki 300-0394, Japan
##Tokai University, Hiratsuka, Kanagawa 259-1292, Japan
***University of Adelaide, SA 5005, Australia
†††Osaka City University, Osaka, Osaka 558-8585, Japan.

**Protons with energies up to ~ $10^{15}$ eV are the main component[1] of cosmic rays, but evidence for the specific locations where they could have been accelerated to these energies has been lacking[2]. Electrons are known to be accelerated to cosmic-ray energies in supernova remnants[3,4], and the shock waves associated with such remnants, when they hit the surrounding interstellar medium, could also provide the energy to accelerate protons. The signature of such a process would be the decay of pions ($\pi^0$), which are generated when the protons collide with atoms and molecules in an interstellar cloud: pion decay results in γ-rays with a particular spectral-energy distribution[5,6]. Here we report the observation of cascade showers of optical photons resulting from γ-rays at energies of ~ $10^{12}$ eV hitting Earth's upper atmosphere, in the direction of the supernova remnant RX J1713.7-3946. The spectrum is a good match to that predicted by pion decay, and cannot be explained by other mechanisms.**

RX J1713.7-3946 is a shell-type supernova remnant (SNR) that was found in the ROSAT all sky survey[7]. Observations with the ASCA satellite have revealed intense non-thermal X-ray emission from the northwest rim[8], TeV γ-ray emission (1 TeV=$10^{12}$eV) has been detected from the northwest rim by the CANGAROO ('collaboration of Australia and Nippon for a γ-ray Observatory in the outback') 3.8-m Cerenkov telescope[9]. Infrared and radio observations have revealed the possible association of the SNR with a molecular cloud complex[10].

The CANGAROO air Cerenkov telescope, which is intended to detect very high energy γ-rays, is located near Woomera, South Australia. The 3.8-m telescope[9], which operated from 1994 to 1998, was replaced in 2000 by a 10-m reflector with a 552-pixel camera of 0.115° square photomultiplier pixels. Observations of RX J1713.7-3946 were carried out 23 - 26 July and 19 - 27 August 2000, and 20 May – 26 June 2001 with the 10-m telescope. After selecting data taken at high elevation angles (>60°) in good weather conditions, a total of 2,332 min on-source and 1,789 min off-source data remained for further analysis.

The differential fluxes of γ-rays from RX J1713.7-3946 are plotted in Fig. 1. The number of excess events was determined from the plots of image orientation angle (α; ref. 11) for on- and off-source runs (Fig. 1 inset) to be 3417±240 (14.3σ). The best fit of the 10-m telescope results gives:

$$dF/dE = (1.63 \pm 0.15 \pm 0.32) \times 10^{-11} \, E^{-2.84 \pm 0.15 \pm 0.20} \quad (1)$$

where the first errors are statistical and the second are systematic with a $\chi^2$/DOF (number degree of freedom) of 0.97, and d$F$/d$E$ is in units of cm$^{-2}$ s$^{-1}$ TeV$^{-1}$. $F$ and $E$ are the gamma-ray flux and energy, respectively. When the results were fitted with a spectrum with an energy cut-off such as $E^{-2}e^{-E/E_{max}}$, an $E_{max}$ of 3.3±0.7 TeV was obtained with $\chi^2$/DOF =2.3. The fluxes are also shown in Fig. 2, multiplied by $E^2$, together with multiwavelength spectra and theoretical predictions (described below). RX J1713.7-3946 is one of the intrinsically brightest galactic TeV γ-ray sources discovered, and it is notable that the power-law spectrum increases monotonically as energy decreases. This is in contrast to the spectrum of SN1006, which flattens below 1 TeV (ref. 12), consistent with synchrotron/inverse Compton (IC) models[13-16]. Both SNRs emit X-rays via the synchrotron process, but the differing TeV spectra suggest a different emission mechanism should be acting in RX J1713.7-3946 at TeV energies.

The morphology of the TeV γ-ray emitting region is shown by the thick solid contours in Fig. 3, together with the ASCA > 2-keV intensity contours[17] and IRAS 100-μm results. The observed TeV γ-ray intensity peak coincides with the northwest rim, and the emission extends over the ASCA contours. A possible extension towards the CO cloud in the northeast[18] can also be seen.

The broadband spectrum plotted in Fig. 2 was derived using data from ATCA (Australia Telescope Compact Array)[19], ASCA[8, 17], the EGRET[20] satellite-borne telescope and this work. In order to explain this spectrum, we considered three mechanisms: the synchrotron/IC process, synchrotron/bremsstrahlung, and $\pi^0$ decay produced by proton-nucleon collisions. The momentum spectra of incident particles (electrons and protons) are assumed to be:

$$dN/cdp = N_0(p/mc)^{-\alpha} \exp(-p/p_{max}) \quad (2)$$

where d$N$/c$dp$ is in units of cm$^{-3}$ TeV$^{-1}$. $N$ is the number of particles, $N_0$ is the normalization factor, $p$ is the momentum of the particle, $m$ is the mass of the particle, $c$ is the velocity of light, $p_{max}$ is the maximum momentum of accelerated particles, and α is the index of the power law spectrum. The effects of acceleration limits from the age and size of the SNR are included in the exponential term. Best-fit values for the radio and X-ray fluxes due to synchrotron radiation from electrons[21] are 2.08 for α, 126 for $p_{max}cB^{0.5}$, and 2.00 for $N_0(V/4\pi d^2)(B)^{(\alpha+1)/2}$; here $V$ is the volume of the radiation region, $d$ is the distance from the Earth, $p_{max}c$ is in units of TeV, and $B$ is in units of μG. The acceptable ranges for these parameters are 1.99—2.13, 93—139 and 4.06—0.58 respectively. The resultant best fit is plotted in Fig. 2 as the solid line.

We initially assumed the 2.7 K cosmic microwave background as the seed photons for IC scattering, where the Klein-Nishina formula was used. A similar model, applied to SN1006, matched the multiband spectrum well[15, 16]. Here, the best-fit values derived for synchrotron radiation from the incident electrons were used, and we assumed that the synchrotron and IC emission regions were the same. Two representative magnetic field strengths, 3 and 10μG, were used. The results, plotted with dotted lines in Fig. 2, show that these models are not consistent with the observed sub-TeV spectrum.

A clear feature of the synchrotron/IC process is the correlation between the peak flux and its energy. We carried out a further study of the above parameter space, taking into account the following uncertainties. For the incident-electron flux, the (1σ) fitting uncertainties were used, and we included IR emission for the IC seed photons. The IR background in the inner region of the Galaxy is not well known. We adopted a model[22] that assumes a temperature of 40 K and an energy density incorporating a dilution factor. The maximum energy density was set to 1eVcm$^{-3}$ in order not to exceed the energy densities of cosmic rays or magnetic fields. For example, IR density at 20' apart from the nearby HII cloud (G347.61+0.20 which is the nearest and brightest IR source) was estimated to be 0.93 eVcm$^{-3}$ using equation (3) in ref. 23. The IC calculation was carried out both with and without this IR maximum energy emission in the IC process, assuming the various incident electron spectra. The shaded area on the right side of Fig. 4 is the resulting theoretically allowed region, with the experimentally allowed region shown by the shaded area on the left side of the figure. The predictions of synchrotron/IC models are inconsistent with our experimental data by an order of magnitude.

Although we have investigated simple cases, there is the possibility that the electron spectrum may soften at higher energies owing to synchrotron cooling. The cooling is severest in the case of the > 10,000-year-old SNR for electrons exceeding $E_{cool}$, which is the critical energy estimated in comparison with the acceleration rate and cooling rate. If $E_{cool} < p_{max}c$, plateau-like features would appear in both the synchrotron and IC fluxes shown in Fig. 2, similar to the cases of pulsar nebulae[24] and blazers[25]. These still would not fit our experimental results.

The bremsstrahlung spectrum was calculated assuming that it occurs in the same region as the synchrotron radiation. A material density of 300 protons cm$^{-3}$ was assumed. The dashed lines in Fig. 2, for magnetic fields of 3 and 10μG, are both inconsistent with our observation. In addition, the bremsstrahlung model is unable to simultaneously reproduce the observed GeV and TeV fluxes. Even if we neglect the EGRET data, a magnetic field of the order of 1G with a material density of 17,500 protons cm$^{-3}$ would be necessary. Both values are unreasonably high.

Thus electron-based models fail to explain the observational results, and so we examined $\pi^0$ decay models. The $\pi^0$s are produced in collisions of accelerated protons with interstellar matter. A model[6] adopting Δ-resonance and scaling was used. We adopted parameters for equation (2) of α = 2.08 and $p_{max}c$ = 10 TeV, considering the plausible parameter regions of typical shock acceleration theory. (If we simply assume the same cut-off energy for electrons, it follows from the best-fit value for the parameter $p_{max}cB^{0.5}$ of 126 that the magnetic field might be as large as 100μG.) The result is shown by the short-long dashed curve in Fig. 2. The best-fit parameters for the total energy of accelerated protons $E_0$ and matter density $n_0$ must satisfy the equation $(E_0/10^{50})(n_0)(d/6)^{-2} = 300$; here $d$ is the distance (in kpc) to RX J1713.7-3946, $E_0$ is in erg, $n_0$ is in protons cm$^{-3}$. A value of $E_0 \approx 10^{50}$ erg gives $n_0$ of the order of 10 or 100 protons cm$^{-3}$ for distances of 1 or 6 kpc, respectively. Both cases are consistent with the molecular column density estimated from Fig. 7 of ref. 10. Assuming a larger value of $p_{max}$ would allow the value of $(E_0/10^{50})(n_0)(d/6)^{-2}$ to be reduced. The $\pi^0$ model alone, therefore, readily explains our results, which provide observational evidence that protons are accelerated in SNRs to at least TeV energies.

We thank H. Tomida for help in understanding the ASCA observational results, M. Seta for comments on molecular cloud density, and F. Aharonian for suggestions. This work was supported by a Grant-in-Aid for Scientific Research by the Japan Ministry of Education, Science, Sports and Culture, and the Australian Research Council.

**Correspondence and requests for materials should be addressed to R.E. (e-mail: enomoto@icrr.u-tokyo.ac.jp).**

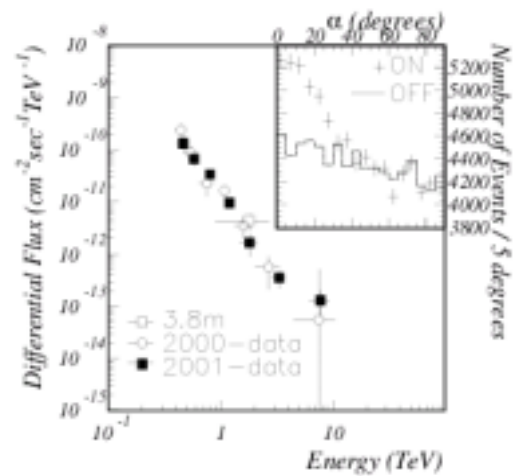

Fig. 1. Differential fluxes, and distributions of the image orientation angle $\alpha$ (inset). Main figure: data shown by open circles and filled squares were obtained in this experiment, the open square shows data from CANGAROO-I (3.8m)[9]. The CANGAROO-I integral flux was multiplied by 1.5 and divided by energy (TeV), assuming that the energy spectrum is proportional to $E^{2.5}$. Inset,

points with error bars show the distribution of a for on-source data, and that for off-source is shown by the histogram. These were obtained as follows: first, 'cleaning' cuts on camera images were applied, requiring pixel pulse-heights of greater than ~3.3 photoelectrons, Cerenkov photon arrival times within 40 ns, and clusters of at least five adjacent triggered pixels in each event. The conventional imaging parameters parameters[11] were then calculated. After applying an energy dependent distance cut, we used the shape parameters to calculate likelihoods *(L;* ref. *26)* under the assumptions of both γ-ray and proton origins for the cascade. The final on- and off-source data sets were obtained by excluding events with a likelihood ratio $L(\gamma\text{-}ray)/\{L(\gamma\text{-}ray)+L(proton)\}$ < 0.4(ref. 27). The number of excess events was determined from the plots of image orientation angle a for on- and off-source runs (inset) to be 3,417±240 (14.3σ). The γ-ray acceptance efficiency was calculated using the Monte Carlo method[26]. In order to check our simulations, we analysed Crab nebula data obtained in December 1999 and December 2000, and found that the flux was consistent with previous experiments[28] to within 12%. In addition, we checked the cosmic-ray spectrum between 100 GeV and 30 TeV. The effects of night-sky background, trigger, and electronics saturation were estimated and included in uncertainties. The systematic uncertainty due to likelihood, clustering methods and pixel triggering threshold was estimated to be 20%. The energy scale uncertainty from the mirror reflectivity, mirror segment distortions, and Mie scatterings was calculated to be 15% (point to point) and 20% (overall).

observation was made with ATCA[19]. The ATCA flux at 1.36 GHz is estimated from two bright filaments lying in the northwest rim of RX J1713.7 - 3946 to be *S* = 4 ± 1 Jy (ref. 19). The shaded area between the thick lines indicates the X-ray emission measured by the GIS detector on the ASCA satellite[17]. The integral flux between 0.5 and 10.0 keV was obtained from Table 4.5 in ref. 17 and the spectral index in Table 4.4. The differential flux was calculated from these two values at 3 keV, the mean GIS sensitivity. The flux uncertainty due to the extended structure of the source was considered to be within +10%/-30%, which was calculated following the procedure described in ref. 29. The EGRET upper limit corresponds to the flux of 3EG J1714-3857[20]. The TeV γ-ray points are from this work (CANGAROO). Lines show model calculations: synchrotron emission (solid line), inverse Compton emission (dotted lines), bremsstrahlung (dashed lines) and emission from $\pi^0$ decay (short-long dashed line). Inverse Compton emission and bremsstrahlung are plotted for two cases: 3 μG (upper curves) and 10 μG (lower curves). The distance to this SNR has ambiguity as follows; the rotation velocity of the associated molecular cloud from this observation yielded a distance to the SNR of 6 ± 1 kpc, in contrast to the distance of 1 kpc estimated from soft-X-ray absorption[8]. The age for the SNR is estimated to be more than 10,000 yr (for a distance of 6 kpc) or ~2,000 yr (for 1 kpc). Details of these models are given in the text.

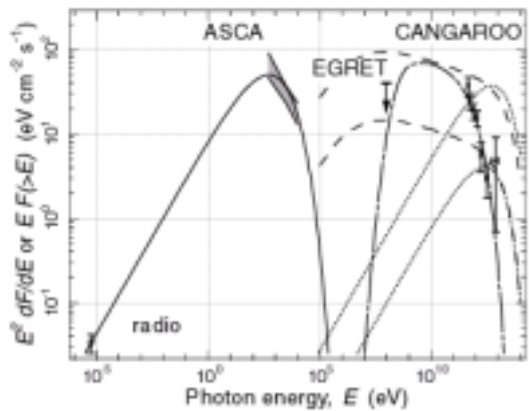

Fig. 2. Multi-band emission from RX J1713.7 - 3946, and emission models. The radio

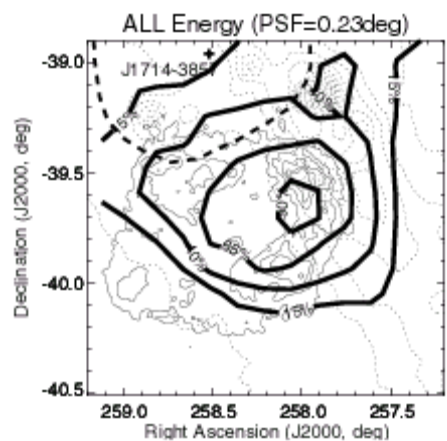

Fig. 3. Profile of emission around the northwest rim of RX J1713.7-3946. The telescope tracked the point in the northwest rim of the SNR from which the maximum X-

ray flux was detected (right ascension 257.9875°, declination-39.5311° in J2000 coordinates)[8]. The solid thick contours are obtained from our observations. These contours were calculated from the distribution of the detection significance determined at each location from the differences in the $\alpha$ plots (on- minus off-source histogram) divided by the statistical errors. The angular resolution was estimated to be 0.23° (1$\sigma$, 68% confidence level). There is a possible systematic offset of the order of 0.1. The detection efficiency drops rapidly outside a 0.5° circle from the northwest rim, that is, the telescope pointing position. The solid thin contours are the ASCA data, and the dotted contours are IRAS 100 $\mu$m data. The sub-GeV source 3EG J1714-3857 is listed in the third EGRET catalogue[20] with a 95% confidence contour which includes the northeast rim (the thick dashed contour). 3EG J1714-3857 is marginally coincident with RX J1713.7-3946 and is an extended source, although there may be source confusion in this area[20].

is defined. On the other hand, theoretical estimates were carried out considering both the 2.7 K CMB alone, and a combination of the IR (40 K) background and the 2.7 K CMB, as target photon sources. The uncertainties from the synchrotron model fitting and those for IR (40 K) flux are taken into account. The shaded area on the right of the figure is the theoretically allowed region (1-$\sigma$) under the assumption of IC process. As there is no overlap with the observationally allowed region, we can rule out IC models for the TeV $\gamma$-ray production.

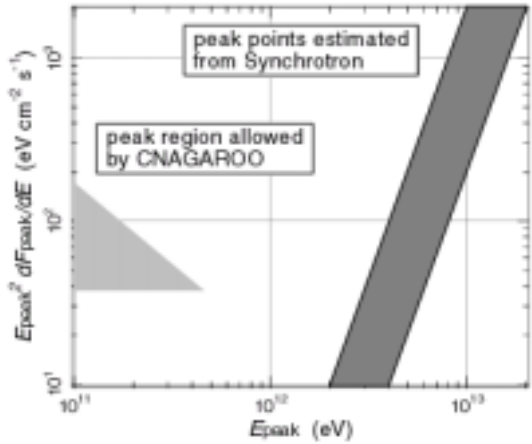

Fig. 4. Allowed regions in the parameter space defined by peak flux ($E^2 dF/dE$) plotted against peak energy. The peak energies and fluxes for IC processes are allowed in this parameter space. The experimental flux (Fig. 2) shows the simple power-law spectrum between 400 GeV and 8 TeV, and hence the peak energy if the IC-process is should be lower than the lowest-energy CANGAROO detection point. This corresponds to the lower-right corner of the shaded area on the left. The energy spectrum below the IC peak must flatten, and so by assuming a power-law spectral index of 2.84 for energies below 400 GeV, the experimentally allowed region